\documentclass[conference]{IEEEtran}
\IEEEoverridecommandlockouts
\usepackage{main}
\usepackage{acronym}
\usepackage{booktabs}

\begin{document}
\begin{acronym}
    \acro{LCDN}{Low-Cost Deterministic Networking}
    \acro{AP}{access point}
    \acro{PTP}[PTP]{Precision Time Protocol}
    \acro{OWD}{One-way delay}
    \acro{RTT}{Round-trip time}
    \acro{SMS}{Smart manufacturing systems}
    \acro{CPS}{Cyber-physical systems}
    \acro{IIoT}{Industrial Internet-of-Things}
    \acro{TUB}{Time Uncertainty Bound}
    \acro{UTC}{Coordinated Universal Time}
    \acro{GPS}{Global Positioning System}
    \acro{OWD}{One-way delay}
    \acro{ppm}{parts per million}
    \acro{RTT}{Round trip time}
    \acro{NIC}{Network interface controllers}
    \acro{DES}{Discrete event simulator}
    \acro{MAC}{media access control}
    \acro{TSN}{Time Sensitive Networking}
    \acro{NTP}{Network Time Protocol}
    \acro{COTS}{commercial off-the-shelf}
    \acro{PDV}{packet delay variation}
    \acro{LAN}{local area networks}
    \acro{SNMP}{Simple Network Management Protocol}
    \acro{SDN}{Software-defined networking}
    \acro{NIC}{Network Interface Card}
    \acro{NC}{Network Calculus}
    \acro{MDR}{Maximum Drift Rate}
    \acro{TDD}{Time Division Duplex}
    \acro{NR}{New Radio}
    \acro{PHC}{PTP Hardware Clock}
    \acro{TT}{TimeTether}
    \acro{API}{Application Programming Interface}
    \acro{DNC}{Deterministic Network Calculus}
    \acro{SP}{strict-priority}
    \acro{IWRR}{Interleaved Weighted Round-Robin}
    \acro{RB}{resource blocks}
    \acro{UE}{User Equipment}
    \acro{OFDM}{Orthogonal Frequency Division Multiplexing}
    \acro{SRS}{Sounding Reference Signal}
    \acro{UL}{Uplink}
    \acro{RRC}{Radio Resource Control}
    \acro{IoT}{Internet of Things}
    \acro{QoS}{Quality of Service}
    \acro{LCDN}{Low-Cost Deterministic Networking}
    \acro{E2E}{End-to-End}
    \acro{JRS}{Joint Routing and Scheduling}
    \acro{TAS}{Time-Aware Shaper}
    \acro{CBS}{Credit-Based Shaper}
    \acro{SPQ}{Strict Priority}
    \acro{ATS}{Asynchronous Traffic Shaper}
    \acro{TC}{Traffic Control}
    \acro{ILP}{Integer Linear Programming}
    \acro{IWRR}{Interleaved Weighted Round Robin}
    \acro{SRP}{Stream Reservation Protocol}
    \acro{RAP}{Resource Allocation Protocol}
    \acro{SOTA}{state-of-the-art}
    \acro{LLDP}{Link Layer Discovery Protocol}
    \acro{PCP}{Priority Code Point}
    \acro{STP}{Spanning Tree Protocol}
\end{acronym}

\newcommand{\gnba}{\ensuremath{\text{gNB}_{\textit{ag}}} }
\newcommand{\gnbi}{\ensuremath{\text{gNB}_{\textit{in}}} }
\newcommand{\ue}[2]{\ensuremath{\text{UE}_{#1}^{\text{#2}}} }
\newcommand{\gain}[2]{\ensuremath{G(#1,#2)} }
\newcommand{\power}[1]{\ensuremath{P_{#1}} }

\title{LCDN: Providing Network Determinism with Low-Cost Switches\\
\author{\IEEEauthorblockN{Philip Diederich, Yash Deshpande, Laura Becker, Wolfgang Kellerer\\
\IEEEauthorblockA{ Chair of Communication Networks, Technical University of Munich, Germany \\
Email: \{philip.diederich, yash.deshpande, laura.alexandra.becker, wolfgang.kellerer\}@tum.de}}
}
}
\maketitle

\begin{abstract}
The demands on networks are increasing at a fast pace. In particular, real-time applications have very strict network requirements. However, building a network that hosts real-time applications is a cost-intensive endeavor, especially for experimental systems such as testbeds. Systems that provide guaranteed real-time networking capabilities usually work with expensive software-defined switches. In contrast, real-time networking systems based on low-cost hardware face the limitation of lower link speeds. This paper fills this gap and presents Low-Cost Deterministic Networking (LCDN), a system designed to work with inexpensive, common off-the-shelf switches and devices. LCDN works at Gigabit speed and enables powerful testbeds to host real-time applications with strict delay guarantees. This paper also provides an evaluation of the determinism of the switch and a Raspberry Pi used as an end device to demonstrate the applicability of LCDN on inexpensive low-power reduced capacity apparatus.

\end{abstract}

\begin{IEEEkeywords}
network performance, Time-sensitive networking. 
\end{IEEEkeywords}

\section{Introduction}
\label{sec:Introdcution}

Emerging applications such as the \ac{IoT}, real-time multimedia, and financial services have increased the need for time-sensitive traffic~\cite{motivationCite, bankingnetworks}. 
These applications require precise guarantees of latency and packet delivery. 
Traditional Ethernet networks, which operate on a best-effort communication model, prioritize network stability and reliability over ensuring real-time data delivery.
IEEE 802.1 \ac{TSN} was developed to tackle these challenges at the data link layer, offering deterministic connectivity and \ac{QoS} within Ethernet networks. 
The TSN task group has established comprehensive standards addressing various methods to ensure latency bounds.

Traditionally, achieving real-time networking capabilities has required specialized and feature-rich switches, such as those supporting \ac{TSN} standards. 
However, these TSN switches' high cost and complexity significantly hinder their widespread adoption in cost-sensitive deployments~\cite{implementationCost} such as testbeds. 
Additionally, many traditional TSN methods exhibit very low network utilization, making running high-bandwidth applications and those requiring deterministic communication challenging. 
As a result, there is a growing need for more affordable and efficient solutions to enable deterministic networking in a broader range of applications.

This paper proposes \ac{LCDN}, a method to provide high-utilization and deterministic data delivery over low-cost switches. 
The motivation for \ac{LCDN} is driven by the increasing feature richness of low-cost switch chipsets, which continue to evolve while maintaining affordability. 
These advancements present an opportunity to implement deterministic networking capabilities on more economical hardware. 
Specifically, the \textbf{contribution} of this paper is that it proposes a novel scheme that provides deterministic networking with low-cost switches. 
The paper details the systems architecture and implementation details, as well as the design choices made, considering the complexity at each step. 
The paper shows via measurements that low-cost switches based on mass-produced ASICs are sufficiently deterministic. 
It shows that certain compensation mechanisms can integrate even reduced capability and low-powered end hosts such as Raspberry Pi into LCDN. 

The rest of the paper is organized as follows: first, it presents the background of deterministic networking and the related \ac{SOTA}. 
This review of related work reveals the need to provide deterministic network guarantees over low-cost switches. 
Then, Section~\ref{sec:main_part} outlines the building blocks of \ac{LCDN}.
Section~\ref{sec:impl} details the implementation of LCDN, detailing the complexity and design choices at each step. 
Section~\ref{sec:switch} presents the measurement of switches to determine the input parameters to the \ac{LCDN} control plane. Next, Section \ref{sec:host} shows the end host measurements to show that the lightweight middleware LCDN needs on the end hosts for data plane management is sufficiently robust.
Finally, Section \ref{sec:conclusion} discusses the limitations of the setup before the conclusion.

\section{Background and related work}
\label{sec:background}

The primary goal of deterministic networks is to maximize the number of accepted flows in the network, ensuring efficient resource management while maintaining the \ac{E2E} latency requirements for each flow \footnote{End-to-end latency of the packet refers to the time taken from the network hardware of the source endpoint to the network hardware of the destination endpoint. The latency incurred as the information passes through the endpoint until it reaches the network hardware is not considered.}.
Meeting this goal involves optimizing two main components: routing and scheduling, known as \ac{JRS}. 
In a given network topology comprising nodes and links, the routing solution determines the path for each flow, providing a sequence of nodes from the source to the destination. 
Each egress port at every node features multiple queue levels. 
Every queue level provides different latency guarantees (i.e., worst-case) based on the scheduling method. 
The scheduling solution then assigns each flow to the appropriate queue level. 

One key \ac{TSN} method is the \ac{TAS}, which facilitates time-triggered transmission for scheduled traffic using gate-controlled queues to manage packet transmission.
\ac{TAS} provides precise transmission slots for traffic with stringent timing requirements.
The \ac{ATS} does not schedule traffic based on time but uses a token-bucket filter for each flow to control the traffic flow and decrease its business. 
Both \ac{TAS} and \ac{ATS} require time-synchronization across devices, although their accuracy needs to be much higher for \ac{TAS}~\cite{tas, ats_timesync}. 
\ac{CBS}, \ac{SPQ}, and other per-class scheduling methods with latency guarantees offer more flexibility and better utilization than \ac{TAS} and also do not require time synchronization. 

All these scheduling methods require a centralized resource management algorithm on the control plane to ensure efficient allocation of time and queues at the switches.
The computational complexity at this centralized controller gets quickly infeasible for \ac{TAS} with an increasing number of flows and size of the network~\cite{tassurvey, jrslimitations}.
Moreover, an effective resource management protocol must handle dynamic changes in network conditions, such as varying traffic loads and potential link failures.

Path diversity between two endpoints in the network must be leveraged to improve the flow acceptance rate. 
Routing packets using just the minimum spanning tree or shortest path is inefficient. 
Given a list of flows, their latency requirements, and the network graph, an \ac{ILP} can be formulated to obtain the best route for any flow~\cite{guckservey}. 
However, the runtime for the solvers becomes quickly intractable~\cite{jrslimitations}.
\ac{TAS} and \ac{ATS} require recomputation when a new flow is added or removed from the network and hence struggle under frequent reconfigurations. 
\ac{TAS} also demonstrates poor network utilization for sporadic traffic~\cite{tassurvey, tas_ats_comparison}.
Chameleon~\cite{van_bemten_chameleon_2020} and Loko~\cite{amaury_loko} use \ac{SPQ} and greedy routing algorithms to achieve fast reconfigurations and high utilization with network determinism. 
However, they require \textit{OpenFlow} \cite{openflow} based \ac{SDN} switches. 
Since Chameleon is intended for data center networks, it uses expensive rack mount switches, while Loko uses the Zodiac FX, which provides only a 100 Mbps rate on each of its four ports. 
Besides, the Zodiac FX switches cost more than the low-cost ones used in \ac{LCDN}. 
Finally, one can drastically reduce the \ac{JRS} problem complexity by using distributed \ac{TSN} protocols such as \ac{SRP} and \ac{RAP}. 
However, due to their limited view of the network, they have significantly reduced network utilization and require signaling overhead to deal with reconfiguration.

The distributed and centralized protocols mentioned above require specialized switches that can perform the bespoke tasks they require.
Since these tasks are specialized, complex~\cite{implementationCost} and not required at mass-scale, our survey found no \ac{TSN} switches commercially available under 1000€. 
The more practical per-class scheduling methods, such as \ac{CBS}, \ac{SPQ}, and \ac{IWRR}, are more readily available on low-cost switches. 
However, a \ac{JRS} problem solution for them to achieve deterministic packet delivery is deficient in the \ac{SOTA}. 
The low-cost switches built for general purpose \ac{LAN} cannot perform arbitrary routing. 
Moreover, a deep evaluation of the performance of these switches with respect to their predictability and determinism is also missing in the \ac{SOTA}. 
\ac{LCDN} aims to bridge this gap to show that real-time deterministic network which achieves high utilization can also be achieved using low cost commercial network hardware. 


\section{System Description}
\label{sec:main_part}
\begin{figure}
    \centering
    \includegraphics[width=0.9\columnwidth]{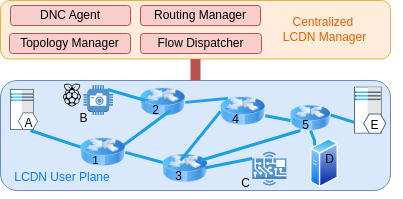}
    \caption{Overview of the LCDN architecture. The switches and end hosts on the user plane connect to the centralized LCDN manager. The manager has four different components for different tasks: the DNC agent checks if a new requested flow can be embedded given the current state of the network by iterating over all paths with the routing agent. It also provides other flows which need to be reconfigured. The topology manager keeps track of the network graph and any connections/disconnections while the flow dispatcher handles communication with the end hosts and reconfigures paths if needed.}
    \label{fig:sys-overview}
\end{figure}

This section introduces the LCDN System and describes its components. 
Figure \ref{fig:sys-overview} shows the conceptual deployment of the LCDN System. There are two main parts: The User Plane and the Centralized LCDN Manager. All the user equipment and switches that comprise the physical network live in the User Plane. 
End-hosts are devices where flows originate or terminate. 
All components in the User Plane are connected to the LCDN Manager, which configures the switches and end-hosts to adhere to the real-time requirements for all the flows in the network.
A lightweight middleware runs on all end-hosts to communicate with the LCDN manager.
This middleware is implemented entirely using the known Linux utility tools, and it remains transparent to the applications generating and receiving the flows.


The LCDN Manager comprises four parts: 1. The Deterministic Network Calculus (DNC) Agent, 2. The Routing Manager, 3. The Topology Manager and 4. The Flow Dispatcher. 
The functionalities provided by these four parts are similar to that of Chameleon \cite{van_bemten_chameleon_2020} and Silo \cite{silo}. 
However, the features provided by low-cost switches differ from these works, hence the LCDN implementation.  
Silo does not utilize the diversity provided by per-class scheduling in the network that LCDN exploits \cite{silo}. 
Chameleon goes one step further by arbitrarily jumping classes at potentially every hop for the same flow~\cite{van_bemten_chameleon_2020} which cannot be implemented in low-cost L2 switches. 
Hence, LCDN solves the JRS problem for a network under the constraint that the priority class of a flow should be preserved from source to destination. 
Moreover, like Chameleon and unlike Silo and QJump, LCDN can reconfigure the path of previously embedded flows when a new flow is requested. 

\textbf{Deterministic Network Calculus Agent:} 
The DNC agent is at the heart of the scheduling part of the JRS in LCDN. 
DNC is a mathematical framework used to analyze the real-time performance of flows in a network~\cite{vanbemtenNetworkCalculusComprehensive2016, le_boudec_network_2001}.
DNC uses models to define traffic patterns and device behavior in the network. 
The models of traffic patterns from data sources are defined in the so-called arrival curves.
Arrival curves describe the amount of data the source produces over time. 
Arrival curves have a constant rate $r$ and a maximum deviation at time $t$ from this rate called burst $b$. 
The device behavior is modeled using the service curves. 
Service curves model the processing times of a switch in different scenarios and the rate at which the switch can forward traffic at a per-class or per-flow level. 
Given a path of a flow and the devices that lie in that path, \ac{DNC} uses min-plus algebra to provide the delay bound, backlog bound, and throughput bound provided that the service curves are strict \cite{le_boudec_network_2001}. 
If the delay bound for a flow is higher than its required specifications, then the DNC agent replies to the routing agent to try another route. 
If the backlog bound exceeds the buffer capacity of any switch or end host in the path, then the flow may experience packet loss, and this flow cannot be embedded in the network, so the routing manager tries a new path. 
Low-cost switches can schedule packets based on their priority according to \ac{SPQ}, WRR, and \ac{IWRR}, for which strict service curves are derived in~\cite{le_boudec_network_2001, vanbemtenNetworkCalculusComprehensive2016, iwrr}. 
To determine the service curve of a switch, one may use the methodology described in Section \ref{sec:switch}, use information from the datasheet, or an RFC2544~\cite{RFC2544} test report if available. 

\textbf{Topology Manager:} 
The Topology Manager builds the Network's topology and translates it into a format the DNC Agent can understand. This task is mainly executed during the set-up phase or when changes in the network occur due to the loss of a link or a device.
Moreover, it maintains $Q\in\mathbb{N}^+$ copies of the topology where $Q$ is the number of priority classes available in the network. 
It then weights each link $i$ with its DNC-provided delay bound for that class.
These weights depend on the arrival curves of embedded flows, the degrees scheduling method, and the parameters of the switch (See Section \ref{sec:switch}.  

\textbf{Routing Manager:} 
The topology manager provides a routing manager with a \textit{queue-level} topology where each link is weighted according to its worst-case delay. 
The routing problem now involves finding a path where the sum of all weights across the path is less than the specified delay bound for that flow.
In LCDN, we assume that flows are embedded one by one. 
It does not need to find the path of all flows simultaneously.
Similar to Chameleon\cite{van_bemten_chameleon_2020}, LCDN uses a greedy routing algorithm for hops. 
It ranks paths according to the sum of their weights in ascending order and embeds the first one that accepts it. 
If all paths are exhausted, it tries to reconfigure already embedded flows. 
While this strategy might have many reconfigurations in the setup phase, but has shown to optimally balance the routing algorithm's time complexity in \cite{van_bemten_chameleon_2020}.

\textbf{Flow Dispatcher:} The Flow Dispatcher is the component that connects to the end hosts. 
A lightweight middleware in the end-hosts ensures that a specified flow's rate and burst parameters are adhered to. 
Secondly, it must tag every packet for a flow with the specified routing and priority values in its header. 
Moreover, to embed a given flow with a specified source and destination, many other end-hosts might need to be contacted if the flows originating from them need to be reconfigured. 
The flow dispatcher does this. 
When an end host wants to send data in the network, it must talk with the Flow Dispatcher via the middleware by sending a flow embedding request which contains the source, destination, rate, burst, and deadline of the flow.
After the DNC Agent and routing manager decide, the Flow Dispatcher either notifies the source that the flow cannot be admitted or sets up the end hosts to send the traffic on a specific route.

\section{Implementation}
\label{sec:impl}
From the above section, it can be seen that switches with specific capabilities are suitable for real-time applications using LCDN. 
These capabilities are (1) A priority-based scheduler at their egress for which a strict service curve is present, (2) They should be manageable to the extent that the topology manager can connect to them and determine the physical network topology, (3) They should be manageable to the extent that the routing manager can route packets arbitrarily in the network. 
We find that almost all COTS-managed switches are capable of this because they are able to (1) Connect to a management interface over telnet, HTTP or \ac{SNMP} (2) VLANs can be assigned to each link, (3) Implement schedulers based on VLAN-priority - also known as \ac{PCP}. 

\textbf{Topology Manager:} 
The setup phase of the network is powered by the \ac{LLDP}, which must be enabled in the switches and the end hosts (via the middleware). 
The topology manager then gets the LLDP report from the devices and builds a graph. 
The routing manager needs the port numbers interconnecting the switches, while the DNC agent needs the maximum link speed for each edge to model the service curve.
Most commercial low-cost switches can process VLAN headers, allowing LCDN to partition the network into different VLANs and route according to a VLAN~\cite{besta_high_performance, otsuka_switch_tagged, kumar_coupling_source}.  
The switches can also run the \ac{STP} per VLAN.
Therefore, LCDN does not have to individually populate the forwarding table of the switches but only manages the multiple spanning trees across all VLANs. 
The routing manager processes the physical network topology from the discovery phase and partitions it into all possible spanning trees using Gabows algorithm~\cite{gabowspanningtrees}, which has a high runtime complexity of $O(V + E + VN)$ where $V$, $E$ and $N$ are the number of vertices, edges and spanning trees respectively. 
This will give $N$ unique paths between all source-destination pairs in the network. 
The VLAN header size restricts the number of VLANs to 4096. However, VLAN ID 0 and VLAN ID 4095 are reserved; hence, if $N > 4094$, LCDN cannot assign a VLAN to that tree. 

To tackle this problem, LCDN first configures only one of the $N$ spanning trees and assigns it VLAN ID 1. 
Then, when the DNC agent accepts a new path that does not belong to the already configured spanning tree(s), the topology manager proceeds to configure that spanning tree. 
This increases the flow configuration time but reduces the overhead of pre-configuring all spanning trees.
This paper leaves the problem of an optimal number of pre-configured spanning trees to reduce the flow configuration time for future work. 
The configuration itself involves a breadth-first search of the spanning tree and assigning the \textit{bridge priority} to each switch ranked according to its depth.  
The \ac{STP} settings allow us to have a maximum of 16 different bridge priorities, placing another limitation on the network size. 
\begin{figure}
    \centering
    \includegraphics[width=0.8\columnwidth]{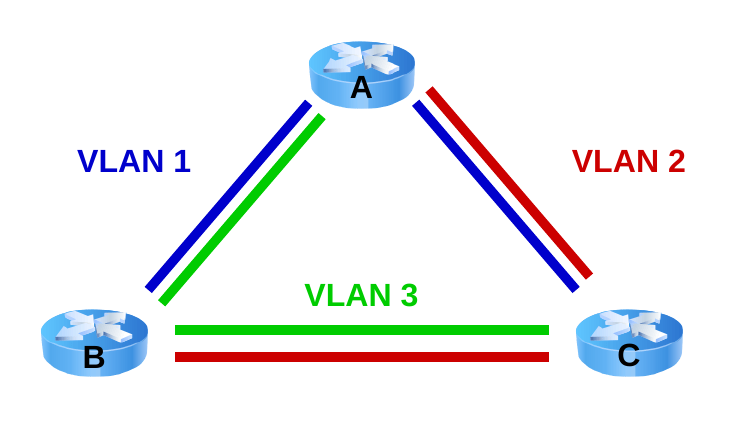}
    \caption{Example assignment of VLAN Identifiers to multiple spanning trees. End-hosts are excluded as they have one connection to a switch and, hence, are already loop-free. 2 VLANs share every single hop path while every 2-hop path has a unique VLAN ID. }
    \label{fig:vlan}
\end{figure}

Figure \ref{fig:vlan} illustrates the idea of VLAN-based routing in a simple scenario with three switches. 
The full mesh network with three switches has three different spanning trees. 
Therefore, in this case, \acs{LCDN} uses a unique VLAN tag for each spanning tree, three (VLANs: 1, 2, 3) and configures the Multiple-STP (MSTP) in switches to select the root node by assigning the higher \textit{bridge priority} to that switch for the VLAN instance of MSTP.
When LCDN routes traffic from switch A to switch C, it can use the blue tree with VLAN ID 1. Whenever the physical resources on the link between switch A and switch C are exhausted, LCDN could route additional traffic from switch A to switch C via the green spanning tree, adding the VLAN tag 3. 
The LCDN middleware tags packets for a particular flow at the source with the corresponding VLAN, which the middleware pops at the destination. 
Therefore, LCDN employs VLAN-based source routing. 
 
\textbf{Routing Manager:}
Like Chameleon~\cite{van_bemten_chameleon_2020}, the routing manager uses Dijkstra over $Q$ trees to find the route, giving this step's complexity as $O(Q(V+E)\log V)$.
The re-routing procedure runs the $k$-shortest path Yen's algorithm~\cite{yen1971finding} over the \textit{physical topology} graph, giving a runtime complexity $O(kV(E + V\log V)$ for each candidate flow that could be rerouted. 
The candidate flows share at least one link with the newly arrived flows selected path.
It is shown in~\cite[Figure 11]{van_bemten_chameleon_2020} that the practical runtime of the rerouting procedure does not increase dramatically compared to the case without rerouting older flows. 
Once a suitable route is found, the topology manager gives its corresponding VLAN ID and also configures a new tree for that VLAN ID if needed. 

\textbf{Miscellaneous:} 
The components that perform the control plane functions of LCDN can be hosted on any end-host in the network. 
It performs in-band management of the switches and end hosts.  
This management traffic is present in the network from the start. 
This management traffic must not be embedded with the highest priority as they are not delay-sensitive. 
Therefore, LCDN embeds the management traffic on VLAN 1 with the lowest priority after the topology discovery phase. 

\textbf{Data Plane:}
The flow admission and access control mechanisms in \acs{LCDN} use network calculus (NC) via the DNC-agent to achieve predictable latencies.
Mathematically, DNC models the traffic sources as arrival curves. 
These arrival curves describe the continuous data rate $r$ and a maximum deviation (burst) $b$ at time $t$.
In the framework, any source must send below the maximum rate and burst advertised in its specifications.
To ensure that the sources in hosts comply with the arrival curves, the \acs{LCDN} uses Token-Bucket Filtering (TBF) from the traffic control \code{tc} utility in Linux's network stack to shape the flows at the source.
The middleware tags the priority and VLAN IDs using VLAN interfaces in Linux.
The middleware is implemented as a RESTful API server to communicate with the control plane functions.

\section{Deterministic Switch}
\label{sec:switch}
\begin{table}[]
    \centering
    \caption{Table summarizes the measured values of the switch. LCDN utilizes these values to model the service curves of the switch using DNC.}
    \begin{tabular}{c c}
       Parameter  &  Value [unit]\\\toprule
       Processing Time  &  4.15 µs \\
       Priority Queues & 8 \\
       Priority Queueing Overhead & 3.5 µs \\
       Buffer Size (per Queue) &  62500 - 500000 Byte
    \end{tabular}
    \label{tab:switch-perf}
\end{table}

\begin{figure}
    \centering
    \includegraphics[width=0.99\columnwidth]{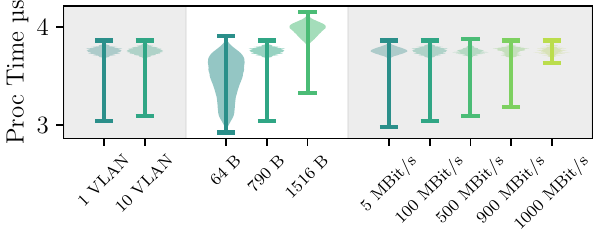}
    \caption{The measured processing times of the switch in various scenarios.}
    \label{fig:proc-times}
\end{figure}

The DNC agent in \acs{LCDN} requires the values for the following parameters for its service curve models: (1) Processing time, (2) Number of Queues at the egress scheduler, (3) Priority Queueing  overhead at the egress scheduler, (4) Buffer Size per Queue. 
In~\cite{amaury_sdn_switches} and~\cite{amaury_loko} the authors provide methods and insights on how to empirically measure determinism in switches. 
The measurements below follow the same methods to determine if the low-cost switches used in LCDN exhibit deterministic behavior. 
Switches with prices as low as 20€, such as TP-Link TL-SG108 and Netgear GS-108, can be used with LCDN. 
Exhaustive measurements in the flowing are carried out on the 80€ FS-S2805S-8TF Switch since we could compare our results with the provided RFC2544 reports from the manufacturer. It has eight RJ-45 ports and two 2 SFP ports. The paper considers only the 8 RJ-45 ports for any measurement.

\textbf{Processing Time:} We vary the number of VLANs, the number of spanning trees, the packet size, and the data rate and measure the processing latency of the switch using the setup from~\cite[Figure 1]{amaury_sdn_switches}. 
Essentially it is an inexpensive method to conduct an RFC2544~\cite{RFC2544} test with taps and a measurement card. 
Figure \ref{fig:proc-times} shows the results for different scenarios. It shows that mainly the packet sizes influence the processing time of the switch. The processing time of the switch behaves deterministic and is upper bounded by \textbf{4.15µs} for the largest packet size of over 10,000 packets.
The number of VLANs and spanning trees, or the load on the network did not affect the processing time of the switch. 

\textbf{Priority Queues:} The number of Queues $Q$, is important for \acs{LCDN} since the Chameleon component uses queue-level graphs instead of link-level graphs to achieve higher path diversity for packets. The switch advertises to have 8 priority queues on the egress side. The paper verifies the number of queues and their order. The switch is set up according to the user manual to achieve strict priority queueing. Then, the second highest priority queue is saturated. To see if \ac{SPQ} works, we send traffic with a higher priority and measure if every packet of the higher priority traffic arrives at the destination. The switch has 8 priority queues.
LCDN can choose to use only a subset of these queues. 

\textbf{Priority Scheduling Overhead:}
 Priority queuing allows higher prioritized traffic to effectively overtake lower priorities within the switch. Whenever the switch sends data out through the egress pipeline, it has to check which priority has data available to send. Then, the switch sends out the data with the highest priority. Checking and deciding which data to send is not for free. Therefore, the switch needs to spend some time calculating from which Queue to send. Chameleon considers this value as well. \acs{LCDN} takes the measurement system from \cite{amaury_sdn_switches} to measure the priority queueing overhead. The measurements and their analysis show that the priority queueing overhead is \textbf{3.5 µs}. The maximum priority queueing overhead appears when a high-priority flow is interrupted by a lower-priority flow with large packet sizes (1516 Bytes). 
 Thus we confirm that the switch adheres to the strict service curve provided for \ac{SPQ} in \cite{le_boudec_network_2001}. 

\textbf{Buffer Size:}
The buffer size is an important parameter for Network Calculus. The bursts of all flows over the same physical port need to fit within the buffer. Otherwise, the system needs to reject potential new flow requests. Manufacturers seldom share information about how the buffer is implemented. Since the switch has a common Realtek Switching Chip, the datasheet presents the total packet buffer capacity, which is 4 Mbit. \acs{LCDN} assumes that the buffer is shared equally between queues and ports. Therefore, the size of the available buffer depends on the number of queues and used ports in the switch. The buffer sizes range from 500000 Byte when only one port is in use to 62500 Byte when all eight ports are in use (for eight priority queues).

Table \ref{tab:switch-perf} summarizes the values \acs{LCDN} uses for the DNC models.

\section{End Host Requirements and Measurements}
\label{sec:host}
\begin{figure}
    \centering
    \begin{subfigure}[b]{0.45\linewidth}
        \centering
        \begin{tikzpicture}
            \begin{axis}[
                xlabel={TBF rate [Mbps]},ylabel={Deviation [\%]},ymin=1.8, ymax=1.9,grid=major,yticklabels={1.80,1.82,1.84,1.86,1.88,1.90},
                xtick={1,2,3,4},xticklabels={1,2,3,4},width=4.4cm, ylabel style={yshift=-0.5em,font=\small},xlabel style={font=\small},xticklabel style={font=\footnotesize},yticklabel style={font=\footnotesize}]
                \addplot[color=blue,mark=square,]
                    coordinates {
                    (0.5,1.8522365759246)
                    (1,1.8523851302829)
                    (1.500, 1.85224721396642)
                    (2.000,1.85178654253861)
                    (2.500,1.85196011375928)
                    (3.000, 1.85069416747454)
                    (3.500,1.84883613252094)
                    (4.000,1.84901738790815)};
            \end{axis}
        \end{tikzpicture}
        \caption{Different Flow Rates}
        \label{fig:rate_host}
    \end{subfigure}
    \hfill
    \begin{subfigure}[b]{0.45\linewidth}
        \centering
        \begin{tikzpicture}
            \begin{axis}[xlabel={TBF Burst size [B]},ylabel={Deviation [\%]},xtick={242, 642, 1042, 1442},grid=major,width=4.5cm,ylabel style={yshift=-1.5em, font=\small},xlabel style={font=\small},xticklabel style={font=\footnotesize, /pgf/number format/1000 sep=},yticklabel style={font=\footnotesize}]
                \addplot[color=red,mark=square,]
                    coordinates {
                    (84,50.00649981552046)
                    (242,13.08895590624894)
                    (442,06.7677969950345)
                    (642,04.56462923611565)
                    (842,03.44413241871526)
                    (1042,02.76546023423554)
                    (1242,02.31063597897005)
                    (1442,01.984468595702)
                    (1542,01.85364426772192)
                    };
            \end{axis}
        \end{tikzpicture}
        \vspace{-0.5cm}
        \caption{Different Burst Sizes}
        \label{fig:graph2}
    \end{subfigure}
    \caption{End host measurements comparing the configured rate of TBF and the measured rate at the receiver. The rate test varies the TBF rate and uses a constant burst size of 1542B (layer 1). The burst size test uses a constant TBF rate of 3Mbps and varies the burst size, which is equal to the application packet size on L1.}
    \vspace{-0.5cm}
    \label{fig:burst_host}
\end{figure}
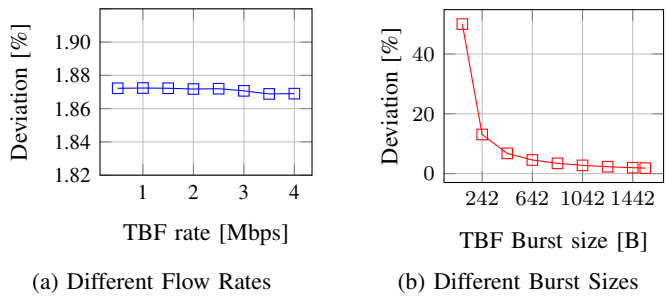

The following validates the TBF's performance in different settings. In the spirit of affordability, the test uses a Raspberry Pi 4 with 8GB of RAM, assuming that any system with more computing resources and superior hardware will perform better. 
The test procedure is as follows: the Raspberry Pi connects directly to a PC with a measurement network card.  Then, the TBF parameters are set on the interface. Next, the Raspberry Pi sends data to the measurement PC. Finally, the measurement PC receives the traffic and calculates the received rate. 
Figure \ref{fig:burst_host} shows the deviation of Token Bucket Parameters and the measured rate in percent. Figure \ref{fig:rate_host} shows the deviation for different rates with a 1542 Byte burst in percent. Here, the result shows that the actual measured rate is slightly higher than the rate set in the TBF, indicating that the filter is leaky. However, the deviation is below 1.88 \% and relatively constant. LCDN can model the deviation of the TBF by adjusting the rate parameter of new flows when sending it to the DNC Agent. 
\ref{fig:graph2} illustrates the deviation of the TBF with different burst sizes and a constant rate of 3\,Mbit/s in percent. Here, the deviation largely depends on the burst size. Furthermore, with small burst values the deviation is large with up to 50 \% deviation. With burst sizes larger than 600 Byte, the deviation is smaller than 5 \%. Even with the large deviation, LCDN can model the behavior of the TBF. For this, LCDN looks at the flow request, checks the burst size, looks up the deviation in a look table, changes the flow rate of the incoming request, and sends it to further processing.


\section{Limitations and Conclusion}
\label{sec:conclusion}

LCDN is designed to work with relatively low-cost hardware. 
In this initial setup, LCDN works with the FS S2805S switch series, which starts at around 80€, and the Raspberry Pi 4, which is in a similar price range. 
Qualitatively 20€ switches also indicate that they could work with LCDN. 
Clearly, these devices have limitations. These limitations, decrease LCDN's performance compared to other TSN methods.
The number of usable VLANs (4094) available on the switches, slower communication with the switch's management interface, and the overhead of configuring multiple spanning trees with constrained hardware all present challenges for using LCDN in small networks or testbeds.   
Moreover, the lack of advanced schedulers like TAS also restrict its use to applications where few time-sensitive flows must co-exist with other background traffic. 

LCDN is a novel system that presents network determinism with affordable hardware. It uses a centralized controller that monitors the whole state of the network.
This allows LCDN to adapt a Network Calculus framework from an already existing controller to suit its use case with low-cost devices. 
Using low-cost and, therefore, lower-powered devices has implications for the system, especially for the routing over L2 switches. LCDN solves the routing problem with multiple spanning trees and source-routed flows. LCDN addresses other integration shortcomings by having additional software components in the data plane that keep track of potential violations. 
That way, LCDN can intervene and preserve network determinism.
This paper also provides important parameters for the NC framework for the FS S2805S switch via measurements. Additionally, this paper measures the determinism of a Raspberry Pi 4 and shows how it can be used with LCDN. 


\section*{Acknowledgements}
This work received funding from Deutsche Forschungsgemeinschaft (DFG, German Research Foundation) - 316878574.
\bibliographystyle{IEEEtran}
\bibliography{references,internal_references}

\end{document}